\documentclass[12pt]{article}

\begin{document}
\vspace*{-.6in}
\thispagestyle{empty}
\begin{flushright}
CALT-68-2078\\
hep-th/9610249
\end{flushright}
\baselineskip = 20pt

\vspace{.5in}
{\Large
\begin{center}
D-Brane Actions with Local Kappa Symmetry\footnote{Work supported in part by
the U.S. Dept. of Energy under Grant No. DE-FG03-92-ER40701.}
\end{center}}

\begin{center}
Mina Aganagic, Costin Popescu, and John H. Schwarz\\
\emph{California Institute of Technology, Pasadena, CA  91125 USA}
\end{center}
\vspace{1in}

\begin{center}
\textbf{Abstract}
\end{center}
\begin{quotation}
\noindent We formulate world-volume actions that describe the dynamics of
Dirichlet $p$-branes in a flat 10d background.  The fields in these theories consist of
the 10d superspace coordinates ($X^m,\theta$) and an abelian world-volume gauge
field $A_\mu$.  The global symmetries are given by the N=2A or N=2B
super-Poincar\'e group, according to whether $p$ is even or odd.
The local symmetries in the $(p + 1)$-dimensional world volume are general coordinate
invariance and a fermionic kappa symmetry.
\end{quotation}
\vfil

\newpage

\pagenumbering{arabic} 

\noindent{\bf Introduction}

During the past year (following the contribution of
Polchinski~\cite{polchinski1}) the important role played by D-branes in
non-perturbative superstring physics has become apparent.  Many of their
remarkable properties have been
elucidated~\cite{witten,polchinski2,danielsson}.  In particular, they have provided a powerful
tool for the study of black holes in string theory~\cite{strominger}.  Very
recently an interesting proposal for understanding non-perturbative 11d (M
theory) physics in terms of ensembles of D 0-branes has been put
forward~\cite{banks}.  For all these reasons it is desirable to achieve as
thorough an understanding of D-branes as possible.  One issue that
has not been explored as thoroughly for D-branes as it has been for more
traditional super $p$-branes (without world-volume gauge fields) is the covariant
formulation of the world-volume action.  A crucial ingredient in such actions
is a local fermionic symmetry of the world volume theory called ``kappa
symmetry.''  It was first identified by Siegel~\cite{siegel} for the
superparticle~\cite{brink}, and subsequently applied to the
superstring~\cite{green1}.  Next it was simplified (to eliminate an unnecessary
vector index) and applied to a super 3-brane in 6d~\cite{hughes}.
Then came the super 2-brane in eleven dimensions~\cite{bergshoeff}, followed by
all super $p$-branes (without world-volume gauge fields)~\cite{achucarro}.

In the case of D-branes, most studies have focused on their bosonic degrees of
freedom and the coupling to bosonic background fields~\cite{douglas,green2}.
Also, some studies have worked in a physical gauge without describing the more
symmetrical gauge-invariant formulas from which they arise.  Interesting as all
of these studies are, there is a subtle and beautiful structure in the
fermionic sector that they do not address.

The main distinction between D-branes and the previously studied super $p$-branes
is that the field content of the world-volume theory includes an abelian vector
gauge field $A_\mu$ in addition to the superspace coordinates $(X^m,\theta)$ of
the ambient $d$-dimensional space-time.  In the case of super $p$-branes whose only
degrees of freedom are $(X^m,\theta),$ $(p + 1)$-dimensional actions have been
formulated that have super-Poincar\'e symmetry in $d$ dimensions realized as a
global symmetry.  In addition they have world-volume general coordinate
invariance, which ensures that only the transverse components of $X^m$ are
physical, and a local fermionic kappa symmetry, which  effectively
eliminates half of the components of $\theta$.  This symmetry reflects the fact
that the presence of the brane breaks half of the supersymmetry in $d$ dimensions,
so that half of it is realized linearly and half of it nonlinearly in the
world-volume theory.  The physical fermions of the world-volume theory
correspond to the Goldstinos associated to the broken supersymmetries.

The purpose of this paper is to present formulas for D-brane actions with local
kappa symmetry analogous to those of the super $p$-branes.  For this purpose the
ambient space-time dimension is restricted to $d = 10$ throughout.  Explicit
Dirichlet $p$-brane actions, with all the appropriate symmetries, will be
presented for all values of $p$ $ (p = 0, 1, \ldots, 9)$.  We know from the rank
of RR gauge fields that when $p$ is even the supersymmetry should be IIA and
when $p$ is odd it should be IIB.  In a physical gauge $X^m$ gives rise to $9-p$
degrees of freedom and $A_\mu$ gives $p - 1$ of them, for a total of 8 bosonic modes.
The 32 $\theta$ coordinates are cut in half by kappa symmetry and in half again
by the equation of motion, so they give rise to 8 fermionic degrees of freedom.

One case, namely $p = 2$, has been studied previously.  As noted in~\cite{duff},
the super 2-brane action in 11d can be converted to the D 2-brane in 10d
by performing a duality transformation in the world volume theory that
replaces one of the $X$ coordinates by a world-volume vector.  This has been
worked out in detail by Townsend~\cite{townsend}, and some of his formulas have
given us guidance in generalizing to all $p$.  (See also~\cite{schmidhuber}.)
One technical detail that aids the analysis is the following:  We do
{\it not} introduce an auxiliary world-volume metric field in the formulas.
They have been included in most studies of super $p$-branes, though this
was not necessary.
If one attempted to incorporate them in the D-brane formulas, this would create
considerable algebraic complications.

\medskip

\noindent{\bf Conventions}

Our conventions are the following.  $X^m$, $m = 0, 1,\ldots, 9$, denotes the 10d
space-time coordinates and $\Gamma^m$ are $32\times 32$ Dirac matrices 
appropriate to 10d with
\begin{equation}
\{\Gamma^m, \Gamma^n\} = 2\eta^{mn}, \quad {\rm where}~ \eta = (- ++ \ldots +).
\end{equation}
(These $\Gamma$'s differ by a factor of $i$ from those of ref.~\cite{green3}.)
The Grassmann coordinates $\theta$ are space-time spinors and world-volume
scalars.  When $p$ is even $\theta$ is Majorana but not Weyl.  It can be
decomposed as
$\theta = \theta_1 + \theta_2$, where
\begin{equation}
\theta_1 = {1 \over 2} (1 + \Gamma_{11})\theta, \qquad \theta_2 = {1 \over 2}
 (1 - \Gamma_{11})\theta.
\end{equation}
These are Majorana--Weyl spinors of opposite chirality.  When
$p$ is odd there are two Majorana--Weyl spinors $\theta_\alpha$ $ (\alpha = 1,2)$
of the same chirality.  The index $\alpha$ will not be displayed explicitly.
The group that naturally acts on it is SL(2,R), whose generators we denote by
Pauli matrices  $\tau_1, \tau_3$.  (We will mostly avoid using $i\tau_2$, which
corresponds to the compact generator.)  With these conventions the
supersymmetry transformations (for all $p$) are given by
\begin{equation}
\delta_\epsilon \theta = \epsilon, \qquad \delta_\epsilon X^m = \bar\epsilon \Gamma^m
\theta.
\end{equation}

World-volume coordinates are denoted $\sigma^\mu, $ $\mu = 0, 1, \ldots, p$.  The
world-volume theory is supposed to have global IIA or IIB super-Poincar\'e
symmetry.  This is achieved by constructing it out of three supersymmetric
quantities.  Besides $\partial_\mu \theta,$ they are
\begin{equation}
\Pi_\mu^m = \partial_\mu X^m - \bar\theta \Gamma^m \partial_\mu \theta,
\end{equation}
and
\begin{equation}
{\cal F}_{\mu\nu} = F_{\mu\nu} - b_{\mu\nu},
\end{equation}
where $F_{\mu\nu} = \partial_\mu A_\nu - \partial_\nu A_\mu$, and $b_{\mu\nu}$
will be defined later.  Another useful quantity is the induced world-volume
metric
\begin{equation}
G_{\mu\nu} = \eta_{mn} \Pi_\mu^m \Pi_\nu^n.
\end{equation}

\medskip

\noindent{\bf The Action}

As in the case of super $p$-branes, the world-volume theory 
of a D-brane is given by a sum of
two terms $S = S_1 + S_2$. The first term
\begin{equation}
S_1 = - \int d^{p + 1} \sigma \sqrt{- {\rm {\rm det}} (G_{\mu\nu} + {\cal F}_{\mu\nu})}
\end{equation}
is essentially an amalgam of the Born--Infeld and Nambu--Goto formulas. The second term
\begin{equation}
S_2 = \int \Omega_{p + 1},
\end{equation}
where $\Omega_{p + 1}$ is a $(p + 1)$-form, is a Wess--Zumino-type term.  $S_1$
and $S_2$ are separately invariant under the global IIA or IIB super-Poincar\'e
group as well as under $(p + 1)$-dimensional general coordinate
transformations.  However, local kappa symmetry will be achieved by a subtle
conspiracy between them, just as in the case of super $p$-branes.

Under local kappa symmetry the variation $\delta\theta$ will be restricted in a
way that will be determined later.  In addition, we require that (whatever
$\delta\theta$ is)
\begin{equation}
\delta X^m = \bar\theta \Gamma^m \delta\theta,
\end{equation}
just as for super $p$-branes.  It follows that
\begin{equation}
\delta\Pi_\mu^m = - 2\delta \bar\theta \Gamma^m \partial_\mu \theta.
\end{equation}
Another useful definition is the ``induced $\gamma$ matrix''
\begin{equation}
\gamma_\mu \equiv \Pi_\mu^m \Gamma_m.
\end{equation}
Note that $\{\gamma_\mu, \gamma_\nu\} = 2 G_{\mu\nu}$.  These formulas imply that
\begin{equation}
\delta G_{\mu\nu} = - 2\delta \bar\theta (\gamma_\mu \partial_\nu + \gamma_\nu
\partial_\mu)\theta.
\end{equation}

The structure of ${\cal F}_{\mu\nu}$ is most easily described in terms of the
2-form ${\cal F} = {1 \over 2}{\cal F}_{\mu\nu} d \sigma^\mu
d\sigma^\nu$.  Then ${\cal F} = F - b$, and the appropriate choice of $b$ turns
out to be~\cite{townsend}
\begin{equation}
b = - \bar\theta \Gamma_{11} \Gamma_m d\theta \left(d X^m + {1\over 2} \bar
\theta \Gamma^m d\theta\right).
\end{equation}
This is the formula for $p$ even.  When $p$ is odd, $\Gamma_{11}$ is replaced by
$\tau_3$.  For this choice
\begin{equation}
\delta_\epsilon b = -\bar\epsilon \Gamma_{11} \Gamma_m d\theta \left(d X^m +
{1\over 2} \bar\theta \Gamma^m d\theta\right)
+ {1 \over 2} \bar\theta \Gamma_{11} \Gamma_m d\theta \bar\epsilon \Gamma^m
d\theta.
\end{equation}
Then $\delta_\epsilon {\cal F} = 0$ if we take
\begin{equation}
\delta_\epsilon A = \bar\epsilon \Gamma_{11} \Gamma_m \theta d X^m + {1\over
6} (\bar\epsilon \Gamma_{11} \Gamma_m \theta \bar\theta \Gamma^m d\theta +
\bar\epsilon \Gamma_m \theta\bar\theta \Gamma_{11} \Gamma^m  d\theta).
\end{equation}
The fundamental identity used to prove this, valid for any three Majorana--Weyl
spinors $\lambda_1, \lambda_2, \lambda_3$ of the same chirality, is
\begin{equation}
\Gamma_m \lambda_1 \bar\lambda_2 \Gamma^m \lambda_3 + \Gamma_m \lambda_2
\bar\lambda_3 \Gamma^m \lambda_1 + \Gamma_m \lambda_3 \bar\lambda_1 \Gamma^m
\lambda_2 = 0. \label{3spinors}
\end{equation}
This formula is valid regardless of whether each of the $\lambda$'s is an even element
or an odd element of the Grassmann algebra.  (Note that $\theta$ is odd and
$d\theta = d \sigma^{\mu} \partial_{\mu} \theta = - \partial_{\mu} \theta d \sigma^{\mu}$ 
is even.)  The variation of ${\cal F}$ under a kappa transformation
is
\begin{equation}
\delta {\cal F} =  2 \delta \bar\theta \Gamma_{11} \Gamma_m d\theta \Pi^m,
\end{equation}
for $p$ even (and $\Gamma_{11} \rightarrow \tau_3$ for $p$ odd) provided that
we decree
\begin{equation}
\delta A = -\delta\bar\theta \Gamma_{11} \Gamma_m \theta \Pi^m + {1\over 2}
\delta \bar\theta \Gamma_{11} \Gamma_{m} \theta \bar \theta \Gamma^m d\theta
- {1\over 2} \delta \bar\theta \Gamma^m \theta \bar\theta \Gamma_{11}
\Gamma_m d\theta.
\end{equation}

\medskip

\noindent{\bf Determination of $S_2$}

Now let's  consider a kappa transformation of $S_1$.  Inserting the variations
$\delta G_{\mu\nu}$ and $\delta {\cal F}_{\mu\nu}$ given above
\[
\delta \left(-\sqrt{-{\rm det} (G + {\cal F})}\right) = - {1\over 2} \sqrt{-{\rm det} (G +
{\cal F})} {\rm tr} [(G + {\cal F})^{-1} (\delta G + \delta {\cal F})]\]
\begin{equation}
= 2 \sqrt{- {\rm det} (G + {\cal F})} \delta \bar\theta \gamma_\mu \{(G - {\cal F}
\Gamma_{11})^{-1}\}^{\mu\nu} \partial_\nu \theta.
\end{equation}
For $p$ odd $\Gamma_{11}$ is replaced this time by $-\tau_3$  (since it has
been moved past $\gamma_{\mu}$).  Now the key step is to rewrite
this in the form
\begin{equation}
\delta L_1 = 2\delta \bar\theta \gamma^{(p)} T_{(p)}^\nu \partial_\nu \theta,
\end{equation}
where
\begin{equation}
\big(\gamma^{(p)}\big)^2 = 1.
\end{equation}
It is not at all obvious that this is possible.  The proof that it is, and the
simultaneous determination of $\gamma^{(p)}$ and $T_{(p)}^\nu$ is the key to
the whole problem.  (The details of the proof will be given elsewhere~\cite{aganagic}.)
Assuming that this is okay, we require that
\begin{equation}
\delta L_2 = 2\delta \bar\theta T^\nu_{(p)} \partial_\nu \theta,
\end{equation}
so that
\begin{equation}
\delta (L_1 + L_2) = 2\delta \bar\theta (1 + \gamma^{(p)}) T^\nu_{(p)} \partial_\nu
\theta.
\end{equation}
Then $\delta \bar\theta = \bar \kappa$, where $\bar \kappa (1 + \gamma^{(p)}) = 0$, gives
the desired symmetry.

It is very useful to define
\begin{equation}
\rho^{(p)} = \sqrt{- {\rm det} (G + {\cal F})} \gamma^{(p)},
\end{equation}
which satisfies
\begin{equation}\label{1star}
\big(\rho^{(p)}\big)^2 = - {\rm det} (G + {\cal F}),
\end{equation}
and to represent it by
\begin{equation}
\rho^{(p)} = {\epsilon^{\mu_{1}\mu_{2}\ldots \mu_{p+1}}\over(p + 1)!}
\rho_{\mu_{1}\mu_{2}\ldots \mu_{p+1}},
\end{equation}
or by a $(p + 1)$-form
\begin{equation}
\rho_{p+1} = {\rho_{\mu_{1}\mu_{2}\ldots \mu_{p+1}}\over (p + 1)!}
d\sigma^{\mu_{1}} d\sigma^{\mu_{2}}\ldots d\sigma^{\mu_{p+1}}.
\end{equation}
The requirement
\begin{equation}
\sqrt{-{\rm det} (G+{\cal F})} \gamma_\mu \{(G - {\cal F}
\Gamma_{11})^{-1}\}^{\mu\nu} = \gamma^{(p)} T_{(p)}^\nu
\end{equation}
can then be recast in the more convenient form
\begin{equation}\label{2star}
\rho^{(p)} \gamma_\mu = T_{(p)}^\nu (G - {\cal F} \Gamma_{11})_{\nu\mu}.
\end{equation}
Writing
\begin{equation}
T_{(p)}^{\nu} = {\epsilon^{\nu_{1}\nu_{2}\ldots\nu_{p} \nu}\over p!}
T_{\nu_{1}\nu_{2}\ldots\nu_{p}},
\end{equation}
$T_{(p)}^\nu$ is characterized by the $p$-form
\begin{equation}
T_p = {T_{\nu_{1}\nu_{2}\ldots\nu_{p}}\over p!} d\sigma^{\nu_{1}}
\ldots d\sigma^{\nu_{p}}.
\end{equation}
In this notation, the kappa variation of $S_2$ takes the form
\begin{equation}
\delta S_2 = 2 (-1)^{p+1}\int \delta \bar\theta T_p d\theta = \delta \int \Omega_{p+1}.
\end{equation}
It is convenient to characterize $S_2$ by a $(p + 2)$-form $I_{p+2} =
d\Omega_{p+1}$.  The preceding formula implies that
\begin{equation}
I_{p+2} = (-1)^{p+1}d\bar\theta T_p d\theta,
\end{equation}
provided that we can show that
\begin{equation}\label{3star}
d\bar\theta\delta T_p d\theta + 2 \delta \bar\theta d T_p d\theta = 0.
\end{equation}
A corollary of
this identity is the closure condition $d I_{p+2} = d \bar\theta dT_p d\theta =0.$

Let us now present the solution of eqs.~(\ref{1star}) and~(\ref{2star}) first
for the case of $p$ even.  For this purpose we define the matrix-valued
one-form
\begin{equation}
\psi \equiv \gamma_\mu d\sigma^\mu = \Pi^m \Gamma_m,
\end{equation}
and introduce the following formal sums of differential forms (the subscript $A$ denotes IIA)
\begin{equation}
\rho_A = \sum_{p={\rm even}} \rho_{p+1}
 \quad {\rm and} \quad T_A = \sum_{p={\rm even}} T_p.
\end{equation}
Then the solution of eqs.~(\ref{1star}) and~(\ref{2star}) is described by the
formulas
\begin{equation}
\rho_A = e^{{\cal F}} S_A (\psi) \quad {\rm and} \quad
T_A = e^{{\cal F}} C_A (\psi)
\end{equation}
where
\begin{equation}
S_A (\psi) = \Gamma_{11} \psi + {1\over 3!} \psi^3 + {1\over 5!} \Gamma_{11}
\psi^5 + {1\over 7!} \psi^7 + \ldots
\end{equation}
\begin{equation}
C_A (\psi) = \Gamma_{11} + {1\over 2!} \psi^2 + {1\over 4!} \Gamma_{11} \psi^4
+ {1\over 6!} \psi^6 + \ldots.
\end{equation}
Thus, $\rho_1 = \Gamma_{11} \psi, $ $\rho_3 = {1\over 6} \psi^3 + {\cal F}
\Gamma_{11} \psi$, etc.,
and $T_0 = \Gamma_{11}, $ $T_2 = {1\over 2} \psi^2 + {\cal F} \Gamma_{11}$, etc.
The fact that $T_0 \not= 0$ means that $S_2 \not= 0$ for the D 0-brane.  The
significant difference from the superparticle of ~\cite{brink} is that the
D 0-brane is massive, whereas the superparticle was massless.

The proof that these expressions for $\rho_A$ and $T_A$ satisfy eq. (\ref{2star}) 
uses the fact that the two terms on the right-hand side of 
the equation correspond to ${1\over 2} \{ \rho^{(p)}, \gamma_{\mu}\}$
and ${1\over 2} [ \rho^{(p)} , \gamma_{\mu}]$.  Using eq.(\ref{3spinors}) one can
also show that this expression for $T_A$ satisfies eq. (\ref{3star}).

The solution for $p$ odd is very similar.  In this case we define (the subscript $B$ denotes IIB)
\begin{equation}
\rho_B = \sum_{p={\rm odd}} \rho_{p+1}\quad {\rm and} \quad
T_B = \sum_{p={\rm odd}} T_p.
\end{equation}
The solution is given by
\begin{equation}
\rho_B = e^{{\cal F}} C_B (\psi)\tau_1 \quad {\rm and} \quad
T_B = e^{{\cal F}} S_B (\psi)\tau_1 ,
\end{equation}
where
\begin{equation}
S_B (\psi) = \psi + {1\over 3!} \tau_3 \psi^3 + {1\over 5!} \psi^5 + {1\over
7!} \tau_3 \psi^7 +\ldots 
\end{equation}
\begin{equation}
C_B (\psi) = \tau_3 + {1\over 2!}\psi^2 + {1\over 4!} \tau_3 \psi^4 + {1\over
6!} \psi^6 +\ldots.
\end{equation}
Thus $\rho_2 = {1\over 2} \tau_1 \psi^2 + i \tau_2 {\cal F}, $ $\rho_4 = 
{1 \over 24} i \tau_2
\psi^4 + {1 \over 2} \tau_1 {\cal F} \psi^2 + {1 \over 2} i \tau_2 {\cal F}^2$, etc., 
and $T_1
= \tau_1 \psi, $ $T_3 = {1 \over 6} i \tau_2 \psi^3 + \tau_1 {\cal F} \psi$, etc.  The
quantity $\rho_0 = i \tau_2$ may be relevant to the D-instanton, which we are
not considering here.

\medskip

\noindent{\bf Conclusion}

This brief letter has not presented proofs of three key identities ---
eqs.~(\ref{1star}),~(\ref{2star}) and~(\ref{3star}).  We plan to write a
longer paper that includes the details of those proofs, as well as some
additional results~\cite{aganagic}.  Other issues to explore
 include the extension of the formulas to non-trivial
space-time backgrounds and the fixing of a physical gauge.  A more challenging
problem is the formulation of generalizations appropriate to the
description of multiple D-branes.

As this work was nearing completion, a paper was posted that gives the kappa
invariant action for the D 3-brane~\cite{cederwall}.  In fact, it also
describes the coupling to background fields.

\end{document}